\documentstyle[twocolumn,prl,aps]{revtex}
\begin{document}
\draft
\title
{Pseudospin Symmetry, Peierls Instability, and Charge Density Wave
in a One-Dimensional Tight-Binding Model
}

\author
{Shun-Qing Shen$^{\dagger}$ and X. C. Xie}

\address
{Department of Physics, Oklahoma State University, Stillwater, OK 74078}
\date{Received July 1995}
\maketitle

\begin{abstract}
We show that there is pseudospin SU(2) symmetry in the one-dimensional
tight-binding model with the inter-molecular
electron-phonon interaction.
We discuss
the relation between the pseudospin
symmetry breaking and the charge-density wave (CDW).
For a finite lattice at half-filling, the Peierls instability
drives the lattice to {\it dimerization} at
low temperature, however, the system remains the SU(2)
symmetry invariant and is not CDW state.
An attractive on-site electron-electron interaction makes the
pseudospin
symmetry spontaneously broken in the thermodynamic limit and the CDW state
arises as well.
Hence, it is clear that spontaneous
pseudospin symmetry breaking produces the CDW state.
\end{abstract}
\pacs{PACS numbers: 74.65+n, 74.20-z}

It was first pointed out by Peierls \cite{Peierls55}
that a one-dimensional metal
coupled to the underlying lattice is unstable at low temperature.
In a half-filling band, the electron-phonon interaction drives
a metal-insulator transition by the dimerization of the lattice due to
Peierls instability. The ground state is a condensate of electron-hole
pairs involving the wave vector $Q=2k_F=\pi/a$ with a non-zero energy
gap.
The condensate is called charge-density wave (CDW) when the electron
density satisfies
\begin{equation}
\rho(n)=1+\rho_Q cos(Qna+\phi)
\end{equation}
where n is the position of the lattice site.
This is a cooperative phenomenon
in a many-body system. In the last four decades, the theory of CDW has
been well established
\cite{Berlinski79,Heeger79,Solyom79,Gruner88,Heeger88}.
The traditional view considering CDW as produced by breaking
of the translational invariance which is a discrete symmetry
in a lattice model. Our present work shows that if the original
model contains the continuous
pseudospin SU(2) symmetry\cite{Jones88,Yang89,Yang90},
then this continuous symmetry has to be broken in order to
have CDW ground state.
Pseudospin SU(2) symmetry is a global
symmetry reflecting the rotation invariance
in the particle-hole
space, while CDW is a condensate of the particle-hole
pairs. Thus, there is a natural connection between the two.
Recently, we discussed the
physical consequences of pseudospin SU(2) symmetry
in the negative-U Hubbard model\cite{Shen94}.
We showed that spontaneous pseudospin SU(2) symmetry breaking will produce
CDW, and furthermore, if its unique subgroup U(1) is also
broken, superconductivity and CDW must coexist to form a supersolid.

CDW is observable in samples with a low dimensional structures,
for example, polyacetylene with a quasi-one-dimensional structure
\cite{Heeger88}.
Successful theories for CDW have been developed
based on the tight-binding models with electron-phonon interaction.
In the tight-binding model there are two dominant sources of the electron
phonon interactions: (i) Inter-molecular or inter-site
(acoustic and optical mode) lattice
phonon modulation of the electronic energies; (ii) Intra-molecular
(or intra-site) vibrational mode modulation of the
electronic energies \cite{Heeger79}.
These two models obey different global symmetries and
play different roles for
CDW occured at low temperatures. In this paper, we first present pseudospin
SU(2) symmetry in a tight-binding model with an inter-molecular
electron-phonon
interaction, then show that the dimerization due to Peierls instability
in this model does not break this SU(2) symmetry and
will not produce CDW.
Introducing an on-site attractive interaction
will lead to the appearance of CDW, and simultaneously pseudospin symmetry
is spontaneously broken in the thermodynamic
limit. The intra-molecular (or intra-site)
electron-phonon interaction breaks pseudospin symmetry explicitly and the CDW
there can be regarded as product of explicit
symmetry breaking.
At the end, we point out that how to distinguish experimentally
the two kinds of CDW resulting from spontaneous and explicit
symmetry breaking.

Our starting point is a simple one-dimensional tight-binding Hamiltonian
with an inter-molecular electron-phonon interaction, i.e.,
the Su-Schrieffer-Heeger (SSH) model \cite{Heeger88},
\begin{equation}
H_{SSH}=H_{\pi}+H^{(i)}_{\pi-ph}+H_{ph},
\end{equation}
 where
\begin{equation}
H_{\pi}=-t_0\sum_{n,\sigma} (c^{\dagger}_{n+1,\sigma} c_{n,\sigma}
+c^{\dagger}_{n,\sigma}c_{n+1,\sigma})
\end{equation}
describes the electron hopping along the one-dimensional chain.
\begin{equation}
H^{(i)}_{\pi-ph}=\alpha\sum_{n,\sigma}(u_{n+1}-u_n)
(c^{\dagger}_{n+1,\sigma} c_{n,\sigma}
+c^{\dagger}_{n,\sigma}c_{n+1,\sigma})
\end{equation}
which is the inter-molecular electron-phonon interaction, and $u_n$ is the
distortion of lattice at site $n$. Finally, the phonon Hamiltonian is written
as
\begin{equation}
H_{ph}=\sum_n \frac{P_n^2}{2M} + \frac{K}{2}\sum_n(u_{n+1}-u_n)^2,
\end{equation}
where $M$ is the mass of the primitive cell and $P_n$ is the momentum
conjugate to $u_n$. In the above model, since
electron hops without flipping its spin,
the total spin ${\bf S}$ is conserved and the system
satisfies the usual spin
SU(2) symmetry. This Hamiltonian is also
invariant under the partial particle-hole transformation\cite{Shiba72},
${\bf T} H_{SSH} {\bf T}^{-1} = H_{SSH}$, where ${\bf T}
=\prod_n(c_{n,\uparrow}-\epsilon(n)c^{\dagger}_{n,\uparrow})$ and $\epsilon(n)
=1$ for even $n$ and $-1$ for odd $n$ \cite{Pu94}.
Combining this symmetry and the spin SU(2)
symmetry gives rise to a {\it hidden} symmetry in the model,
{\it i.e.}, pseudospin SU(2) symmetry. The pseudospin operators are defined by
\begin{equation}
\left\{
\begin{array}{l}
\tilde{\bf S}^+=\sum_n\tilde{\bf S}^+_n=\sum_n(-1)^n c_{n,\uparrow}
c_{n,\downarrow},\\
\tilde{\bf S}^-=\sum_n\tilde{\bf S}^-_n=\sum_n(-1)^n
c^{\dagger}_{n,\downarrow}
c^{\dagger}_{n,\uparrow},\\
\tilde{\bf S}^z=\sum_n\tilde{\bf S}^z_n=\sum_n\frac{1}{2}
(1-c^{\dagger}_{n,\uparrow}c_{n,\uparrow}-
c_{n,\downarrow}^{\dagger}c_{n,\downarrow})
\end{array}
\right.
\end{equation}
which construct a SU(2) group. Denote ${\bf O}
=e^{i\delta {\bf n}\cdot\tilde{{\bf S}}}$ ($\delta$ is arbitrary and {\bf n}
is a unit vector), we have ${\bf O}H_{SSH}{\bf O}^{\dagger}=H_{SSH}$. The
physical consequences related to this symmetry
have been discussed extensively in the literature
\cite{Yang89,Shen94,Zhang90,Shen93}.
One of them
is the possible relevance to the appearance of CDW while the
symmetry is broken \cite{Shen94}.
It is worth of mentioning here that in this paper we only discuss the case
that pseudospin SU(2) symmetry is broken while U(1) subgroup is invariant.

Usually the SSH model is used to explain Peierls
transition and its soliton-like excitation in some one-dimensional materials.
Peierls showed that within the mean field approximation, the ground state
of a one-dimensional metal undergoes a spontaneous lattice distortion
with
$\langle u_n \rangle\neq 0$. This instability occurs at $Q=2k_F=\pi/a$ (a is
the lattice unit length), and the distortion with the mean amplitude,
$u_n\rightarrow \langle u_n \rangle=(-1)^n u$.
In this case, the SSH Hamiltonian
is reduced to
\begin{eqnarray}
H^0(u)&=&-\sum_{n,\sigma}(t_0+2\alpha (-1)^n u)(c^{\dagger}_{n+1,\sigma}c_{n,
\sigma}+c^{\dagger}_{n,\sigma}c_{n+1,\sigma})\nonumber \\
&+&2KN u^2.
\end{eqnarray}
This Hamiltonian can be diagonalized to
\begin{equation}
H^0(u)=-\sum_{k,\sigma}\omega^0_{ k}
(\alpha^{0\dagger}_{k,\sigma}\alpha^0_{ k,\sigma}
-\beta^{0\dagger}_{ k,\sigma}\beta^0_{k,\sigma})
+2KNu^2,
\end{equation}
by the following Bogoliubov transformation
\begin{equation}
\left (
\begin{array}{c} \alpha^0_{ k,\sigma} \\ \beta^0_{k,\sigma}
\end{array} \right )
=\left (
\begin{array}{cc}
\sin \theta^0_{k} & -i\cos \theta^0_{k} \\
-i\cos \theta^0_{k}   & \sin \theta^0_{k} \end{array}
\right ) \left (
\begin{array}{c} c_{k,\sigma} \\ c_{ k+Q,\sigma} \end{array}
\right ),
\end{equation}
where $\omega^0_{k}=(\gamma^2_{k}+\Delta^2_{k})^{1/2}$
($\gamma_{k}=-2t\cos ka$ and $\Delta_k=4\alpha u \sin ka$)
and $\sin \theta^0_k =\frac{1}{\sqrt{2}}(1-\frac{\gamma_k}{\omega^0_k})^{1/2}$,
$\cos \theta^0_k =\frac{1}{\sqrt{2}}(1+\frac{\gamma_k}{\omega^0_k})^{1/2}sgn(
\Delta_k)$ ($-\pi/2a\leq k\leq \pi/2a$). The order parameter $u$ can be
determined self-consistently. When $u\neq 0$, there is a gap,
$8\alpha u$,
between the ground state and the first excitation state in a half-filled band.
Thus, the system becomes an insulator due to the lattice distortion. For
our purpose, we write the
ground state as
\begin{equation}
\vert P \rangle
=\prod_{k,\sigma}(\sin \theta^0_k +i\cos \theta^0_k b^{\dagger}_{k,\sigma})
\vert F\rangle
\end{equation}
where
$b^{\dagger}_{k,\sigma}= c^{\dagger}_{k+Q,\sigma}
 c_{k,\sigma}$ is the creation operator for a particle-hole pair
and $\vert F\rangle
 =\prod_{k,\sigma}c^{\dagger}_{k,\sigma}\vert 0\rangle$ is the ground state
of $H_{\pi}$, {\it i.e.}, the Hamiltonian without any lattice distortions.
The state $\vert P \rangle $ is a
condensate of electron-hole pairs involving the momentum $Q$. Like
the superconducting BCS state Cooper pairs, and the
number of particle-hole pairs is undetermined in $\vert P \rangle $.
The price of this uncertainty is that
the order parameter could have an arbitrary phase factor since
the phase and the
number of the pairs are conjugate to each other and obey the uncertainty
relation. Furthermore,
it is easy to calculate that
\begin{equation}
\tilde{{\bf S}}^{\pm}\vert P\rangle =
\tilde{{\bf S}}^z\vert P\rangle =0,
\end{equation}
which indicates that $\vert P\rangle$ is an eigenstate of pseudospin operators
with zero eigenvalue.
Hence, $\vert P \rangle $
keeps pseudospin
symmetry invariant, namely, rotation invariance in particle-hole space,
${\bf O}\vert P\rangle=\vert P \rangle$. Meanwhile,
the order parameter for CDW is also found to be zero in $\vert P \rangle $,
\begin{eqnarray}
\rho_Q&=&\frac{1}{N}\sum_{k,\sigma}
\langle P\vert c^{\dagger}_{k+Q,\sigma} c_{k,\sigma}+
 c^{\dagger}_{k,\sigma} c_{k+Q,\sigma}\vert P\rangle \nonumber \\
&=& \frac{1}{N}\sum_{k,\sigma}(i \sin\theta^0_k \cos\theta^0_k
-i\sin\theta^0_k\cos\theta^0_k)=0
\end{eqnarray}
Therefore, the dimerization of the lattice
caused by Peierls instability does not break
pseudospin symmetry and will not produce CDW since CDW corresponds to
oscillating electron density wave with nonzero $\rho_Q$.
Due to the invariance of pseudospin symmetry, there is no CDW while the
lattice dimerizes driven by the inter-molecular electron-phonon interaction.
The Hamiltonian in Eq.(7)
satisfies the pseudospin symmetry and is exactly solvable.
Its unique ground state must be invariant under this symmetry.
The corresponding charge density is always uniform, i.e.,
\begin{eqnarray}
\rho(n)&=&\langle P \vert \sum_{\sigma}c^{\dagger}_{n,\sigma} c_{n,\sigma}
\vert P \rangle=
\langle P \vert {\bf O}^{\dagger}
\sum_{\sigma}c^{\dagger}_{n,\sigma}c_{n,\sigma}
{\bf O}\vert P\rangle\nonumber \\
&=&\langle P \vert 2- (\sum_{\sigma}c^{\dagger}_{n,\sigma} c_{n,\sigma}) \vert
P \rangle  =1
\end{eqnarray}
For any finite systems,
the symmetry ensures this uniform density result to hold.
Usually it is thought that breaking of the translational symmetry
(which is discrete in this case) will lead
to CDW. We find that additional continuous pseudospin symmetry
has to be broken to produce CDW.

In order to further
illustrate that CDW is the product of pseudospin symmetry breaking,
we introduce an on-site electron-electron interaction which also possesses
pseudospin symmetry,
\begin{equation}
H_I=U\sum_{n,\sigma}(c^{\dagger}_{n,\uparrow}c_{n,\uparrow}
-\frac{1}{2})(c^{\dagger}_{n,\downarrow}c_{n,\downarrow}-\frac{1}{2}).
\end{equation}
Due to the particle-hole symmetry in the total Hamiltonian
\begin{equation}
H_t=H_{SSH}+H_I,
\end{equation}
the chemical potential is equal to zero at a half filling.
As has shown previously, there
is a condensate of the particle-hole pairs involving the momentum
$Q=2k_F$ in the unperturbed ground state, although it is not a CDW state since
$\rho_Q =0$. This fact reflects the instability of the unperturbed state, which
tends to form a CDW consisting of particle-hole pairs.
After introducing the electron-electron
interaction, whether the condensate is stable is concerned here. Perturbation
calculation with finite orders can not break the symmetry if we
regard the the electron-electron interaction $H_I$ as a perturbation.
It is obvious that a
self-consistent calculation is necessary.
The approach using here is quite similar as the BCS mean field approach for
superconductivity. We assume that for the ground state of $H_t$,
$\langle g \vert
c^{\dagger}_{k,\sigma}c_{k+Q,\sigma}\vert g \rangle\neq 0$.
Under this assumption,
$H_I$ can be decoupled by introducing an order parameter
$\rho$,
\begin{equation}
H_I\rightarrow \rho\sum_{k,\sigma}(c^{\dagger}_{k,\sigma}c_{k+Q,\sigma}+
c^{\dagger}_{k+Q,\sigma}c_{k,\sigma})-\frac{1}{2U}N\rho^2
\end{equation}
and the order parameter
\begin{equation}
\rho=\frac{U}{N}\sum_{k,\sigma}\langle g\vert
c^{\dagger}_{k,\sigma}c_{k+Q,\sigma}+
c^{\dagger}_{k+Q,\sigma}c_{k,\sigma}
\vert g\rangle=U\rho_Q.
\end{equation}
 The Hamiltonian $H_t$ can then be reduced to
\begin{eqnarray}
H_{MF} &=&
\sum_{k,\sigma}\{\gamma_k(
c^{\dagger}_{k,\sigma}c_{k,\sigma}-
c^{\dagger}_{k+Q,\sigma}c_{k+Q,\sigma})\nonumber \\
&+&[(\rho-i\Delta_k)c^{\dagger}_{k+Q,\sigma}c_{k,\sigma}
+(\rho+i\Delta_k)c^{\dagger}_{k,\sigma}c_{k+Q,\sigma}]\}\nonumber \\
&-&\frac{1}{2U}N\rho^2 + 2NKu^2.
\end{eqnarray}
Using the Bogoliubov transformation once more,
\begin{equation}
\left (
\begin{array}{c} \alpha_{ k,\sigma} \\ \beta_{k,\sigma}
\end{array} \right )
=\left (
\begin{array}{cc}
\sin \theta_{k} & -i\cos \theta_{k}e^{-i\phi_k} \\
-i\cos \theta_{k}e^{i\phi_k}   & \sin \theta_{k} \end{array}
\right ) \left (
\begin{array}{c} c_{k,\sigma} \\ c_{ k+Q,\sigma} \end{array}
\right ),
\end{equation}
one gets
\begin{eqnarray}
H_{MF}&=&-\sum_{k,\sigma}\omega_k(\alpha^{\dagger}_{k,\sigma}\alpha_{k,\sigma}
-\beta_{k,\sigma}^{\dagger}\beta_{k,\sigma})
\nonumber \\
&-&\frac{1}{2U}N\rho^2+2NKu^2,
\end{eqnarray}
where $\omega_k=(\gamma_k^2+\rho^2+\Delta_k^2)^{1/2}$,
$\sin \theta_k=\frac{1}{\sqrt{2}}(1-\frac{\gamma_k}{\omega_k})^{1/2}$,
$\cos \theta_k=\frac{1}{\sqrt{2}}(1+\frac{\gamma_k}{\omega_k})^{1/2}
sgn(\Delta_k)$,
and $e^{i\phi_k}=\left (\frac{\Delta_k+i\rho}{\Delta_k-i\rho}\right )^{1/2}$
The
order parameters are determined by a set of equations,
\begin{equation}
\left \{ \begin{array}{l}
\displaystyle\frac{1}{N}\sum_k \displaystyle\frac{1}{\omega_k} =-
\displaystyle\frac{1}{2U},\\
\displaystyle\frac{1}{N}\sum_k \displaystyle\frac{\Delta_k^2}{\omega_k u^2}=2K.
\end{array}
\right.
\end{equation}
If $U>0$, the order parameter $\rho$ is always equal to zero and no
CDW arises. The non-zero $\rho$ and CDW ground state
only exists for attractive interaction with $U<0$. Here
again, the CDW ground state is expressed in the form
\begin{equation}
\vert g\rangle=\prod_{k,\sigma}(
\sin \theta_k +i\cos \theta_k e^{i\phi_k}
b^{\dagger}_{k,\sigma})\vert F\rangle.
\end{equation}
When $\rho=0$, the state $\vert g\rangle=\vert P \rangle$, i.e., the Peierls
state.
It is easy to check that $\tilde{S}^z\vert g\rangle =0$, which indicates
the U(1) symmetry is invariant in the ground state, but after some algebra
one can show
\begin{eqnarray}
\tilde{S}^+\vert g \rangle&=&\sum_{k}(c_{k+Q,\uparrow}c_{-k,\downarrow}+
c_{k,\uparrow}c_{-k-Q,\downarrow})\vert g\rangle \nonumber \\
&=&-i\sum_k(\sin\theta_{-k}\cos\theta_k e^{i\phi_k}+
\sin\theta_k\cos\theta_{-k}e^{i\phi_{-k}})\nonumber \\
&\times&\prod_{k',\sigma\neq k,\uparrow,
-k,\downarrow}(
\sin \theta_{k'}c^{\dagger}_{k',\sigma}+i
\cos \theta_{k'}e^{i\phi_{k'}}
c^{\dagger}_{k'+Q,\sigma})\vert 0\rangle
\nonumber \\
&=& \rho\sum_k\vert g_k\rangle
\end{eqnarray}
where
\begin{equation}
\vert g_k\rangle=-\frac{1}{\omega_k}\prod_{k',\sigma\neq k,\uparrow,
-k,\downarrow}(
\sin \theta_{k'}c^{\dagger}_{k',\sigma}+i
\cos \theta_{k'}e^{i\phi_{k'}}
c^{\dagger}_{k'+Q,\sigma})\vert 0\rangle.
\end{equation}
If $\rho\neq 0$, then $\tilde{S}^+\vert g\rangle\neq 0$. Thus, in this state,
pseudospin symmetry is spontaneously broken as soon as CDW arises. The
appearance of CDW has one-to-one correspondence to the
pseudospin symmetry breaking. In
other words, spontaneous pseudospin symmetry breaking will produce
the CDW, just as superfluid or superconductivity is the product
of the global U(1) symmetry spontaneous breaking.

For both models without or with interaction
defined by Eqs.(2) and (15), it has been rigorously shown that
in finite systems with even number sites their
ground states have pseudospin $\tilde{\bf S}=0$ \cite{Shen93}, therefore,
the pseudospin symmetry is not spontaneously broken. The spontaneous symmetry
breaking can only happen in the thermodynamic limit, similar as in
an antiferromagnet system.
In the case of attractive interaction in Eq.(15), the ground states
might be degenerate in the thermodynamic limit.
The degeneracy makes it possible for the spontaneous pseudospin symmetry
breaking in the ground state. In fact, the mean-field
ground state $\vert g\rangle$ consists of all eigenstates of $\tilde{\bf S}$
and reflects this degeneracy.
Similar physics happens for the BCS ground state
which consists of a series of eigenstates of particle
numbers. The spontaneous symmetry breaking here is different from that in a
ferromagnet system. In that case,
SU(2) spin symmetry could be spontaneously
broken even in a finite system.

Before ending the paper, we should point out that
CDW could also be the product of
explicit pseudospin SU(2) symmetry breaking. As mentioned previously, the
intra-molecular (or intra-site) vibrational mode
modulation of the electronic energies is
another dominant source of the electron-phonon interaction, which could also
lead to the appearance of CDW.
This kind of interactions is expressed as\cite{Heeger79}
\begin{equation}
H^{(ii)}_{e-ph}=\sum_{n,\sigma}(\sum_{\nu}\Gamma_{\nu}u_{\nu,n})
c^{\dagger}_{n,\sigma}c_{n,\sigma},
\end{equation}
where $\Gamma_{\nu}$ represents the coupling of the electron to the mode,
say at $q=Q$.
This interaction breaks pseudospin symmetry explicitly. The condensation
of the phonons make it possible to produce a effective periodic electric
field which drives a CDW. This is just like that a periodic magnetic
field will produce a spin-density wave in a spin system. From the
point of view of symmetry, the physical mechanisms of these two kinds
of CDW are very different. We can distinguish them by measuring the
the low-energy collective modes. According to the Goldstone theorem
\cite{Goldstone60},
spontaneous pseudospin symmetry breaking will produce two massless
collective modes while U(1) symmetry is invariant, and one of them could
be measured experimentally, as we discussed in a recent paper \cite{Shen94}.

In summary, we discuss pseudospin SU(2) symmetry and its physical consequences
in the one-dimensional
tight-binding model with the electron-phonon interaction and on-site
electron-electron
interaction. The one-dimensional lattice dimerization does not
break pseudospin symmetry and will not produce CDW. We
show that on-site attractive interaction will produce CDW combining the
Peierls instability, and simultaneously break pseudospin symmetry in the
ground state. Therefore, we conclude that spontaneous pseudospin SU(2)
symmetry
breaking will produce CDW.


\end{document}